\documentclass[twocolumn]{aastex631}

\usepackage{amsmath,amsthm,amsfonts,latexsym,amscd, tabularx}
\usepackage{graphicx}
\usepackage{longtable}
\usepackage{array}
\usepackage{gensymb}

\graphicspath{ {7027_Figs/} }

\newcommand{\dotarcsec}{\(\stackrel{''}{\textstyle.}\)}

\received{September 13, 2022}
\revised{November 16, 2022}
\accepted{November 18, 2022}

\submitjournal{ApJ}

\begin{document}

\title{Mapping NGC 7027 in New Light: CO$^+$ and HCO$^+$ Emission Reveal Its Photon- and X-ray-Dominated Regions}

\author[0000-0002-5714-7367]{Jesse Bublitz}
\affiliation{Green Bank Observatory, 155 Observatory Road, Green Bank, WV 24944, USA}
\affiliation{Chester F. Carlson Center for Imaging Science and Laboratory for Multiwavelength Astrophysics, Rochester Institute of Technology, 54 Lomb Memorial Drive, Rochester, NY 14623, USA}
\author[0000-0002-3138-8250]{Joel H. Kastner}
\affiliation{Chester F. Carlson Center for Imaging Science and Laboratory for Multiwavelength Astrophysics, Rochester Institute of Technology, 54 Lomb Memorial Drive, Rochester, NY 14623, USA}
\author{Pierre Hily-Blant}
\affiliation{Université Grenoble Alpes, CNRS, IPAG, F-38000 Grenoble, France}
\author[0000-0003-0536-4607]{Thierry Forveille}
\affiliation{Université Grenoble Alpes, CNRS, IPAG, F-38000 Grenoble, France}
\author[0000-0002-7338-0986]{Miguel Santander-Garc\'{i}a}
\affiliation{Observatorio Astron\'{o}mico Nacional, Alfonso XII, 3, 28014, Madrid, Spain}
\author[0000-0003-1968-0117]{Javier Alcolea}
\affiliation{Observatorio Astron\'{o}mico Nacional, Alfonso XII, 3, 28014, Madrid, Spain}
\author[0000-0002-6586-4665]{Valentin Bujarrabal}
\affiliation{Observatorio Astron\'{o}mico Nacional, Alfonso XII, 3, 28014, Madrid, Spain}
\author[0000-0003-1526-7587]{David J. Wilner}
\affiliation{Harvard-Smithsonian Center for Astrophysics, 60 Garden Street, Cambridge, MA 02138, USA}
\author[0000-0002-6752-2909]{Rodolfo Montez, Jr.}
\affiliation{Harvard-Smithsonian Center for Astrophysics, 60 Garden Street, Cambridge, MA 02138, USA}
\author[0000-0002-7989-9041]{Isabel Aleman}
\affiliation{UNIFEI, Instituto de F\'{i}sica e Qu\'{i}mica, Universidade Federal de Itajub\'{a}, Av. BPS 1303 Pinheirinho, 37500-903 Itajub\'{a}, MG, Brazil}



\begin{abstract}

The young and well-studied planetary nebula NGC 7027 harbors significant molecular gas that is irradiated by luminous, point-like UV (central star) and diffuse (shocked nebular) X-ray emission. This nebula represents an excellent subject to investigate the molecular chemistry and physical conditions within photon- and X-ray-dominated regions (PDRs and XDRs). As yet, the exact formation routes of CO$^+$ and HCO$^+$ in PN environments remain uncertain. Here, we present $\sim$2$''$ resolution maps of NGC 7027 in the irradiation tracers CO$^+$ and HCO$^+$, obtained with the IRAM NOEMA interferometer, along with SMA CO and HST 2.12~$\mu$m H$_2$ data for context. The CO$^+$ map constitutes the first interferometric map of this molecular ion in any PN. Comparison of CO$^+$ and HCO$^+$ maps reveal strikingly different emission morphologies, as well as a systematic spatial displacement between the two molecules; the regions of brightest HCO$^+$, found along the central waist of the nebula, are radially offset by $\sim$1$''$ ($\sim$900 au) outside the corresponding CO$^+$ emission peaks. The CO$^+$ emission furthermore precisely traces the inner boundaries of the nebula's PDR (as delineated by near-IR H$_2$ emission), suggesting that central star UV emission drives CO$^+$ formation. The displacement of HCO$^+$ radially outward with respect to CO$^+$ is indicative that dust-penetrating soft X-rays are responsible for enhancing the HCO$^+$ abundance in the surrounding molecular envelope, forming an XDR. These interferometric CO$^+$ and HCO$^+$ observations of NGC 7027 thus clearly establish the spatial distinction between the PDR and XDR formed (respectively) by intense UV and X-ray irradiation of molecular gas.

\end{abstract}



\section{Introduction}

As a nearby \citep[$D \sim 890$ pc;][]{Masson89, Ali15},  young \citep[dynamical age $\sim$1000~yrs;][]{Ali15, Schonberner18} and rapidly evolving planetary nebula (PN) with a massive molecular envelope, NGC 7027 represents an excellent source to study the effects of UV and X-ray-induced chemical processes in molecular gas. 
Photometric studies have identified a hot central star of $T_{eff}$ $\sim$200~kK that drives photoionization and photodissociation in the ejected envelope \citep{Graham93, Latter00, Zhang05, Moraga2022}. The central star (CSPN) of NGC 7027 is a luminous extreme UV source \citep[$L_{UV} \sim 3\times10^{37}$ ergs s$^{-1}$;][]{Latter00,Moraga2022},
while Chandra X-ray Observatory observations established NGC 7027 as among the most luminous, diffuse X-ray sources among PNe, with a luminosity of $L_X$ = 7$\times$10$^{31}$ ergs s$^{-1}$ \citep{Kastner12, Montez18}. Optical and IR observations indicate a carbon-rich nebula \citep[C/O = 2.7;][]{Zhang05}, which is consistent with detections of C- and O-based species in the molecular envelope, suggesting a progenitor mass of 2--4 $M_{\odot}$ \citep[][and references therein]{KastnerWilson2021}. The temperature and luminosity of NGC 7027's CSPN indicate a present-day mass of $\sim$0.7 $M_{\odot}$ \citep{Moraga2022}.


\begin{figure*}
	\centering
	\includegraphics[width=\textwidth]{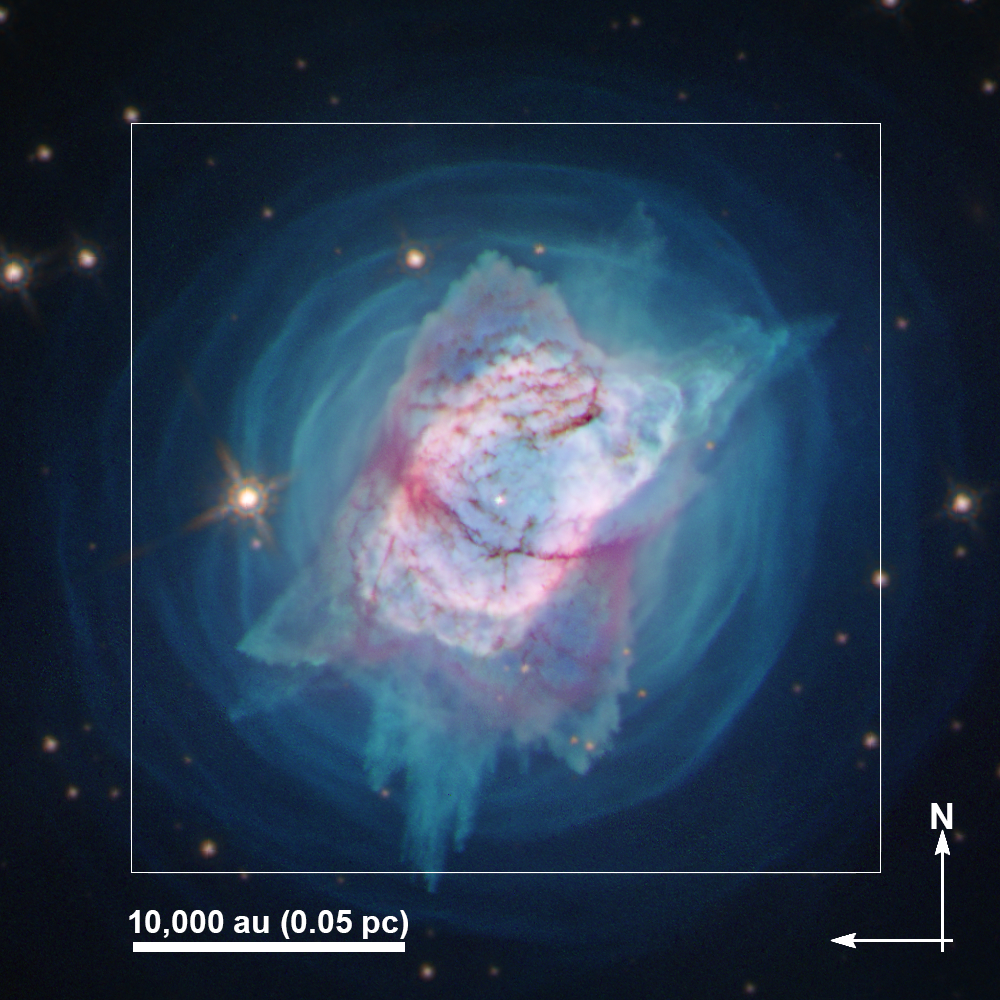}
	\caption{Composite image of NGC 7027 from HST/WFC3 using wide IR (red: F110W, blue: F160W) and narrow band (red: F128N+F130N, green: F673N+F656N, blue: F502N+F342N+F487N) filters. Field of view is 40$'' \times$40$''$. The white box marks the 30$'' \times$30$''$ field of view of other figures in this work. N is up, E is to the left. Image credit: STScI.}
	\label{fig_7027HST}
\end{figure*}

NGC 7027 yielded the first identification of a molecule in any PN \citep[CO;][]{Mufson75} and now boasts one of the largest catalogs of molecular transitions detected in a single PN.
The nebula is particularly rich in emission from molecular ions, including detections of HCO$^+$, CO$^+$, and N$_2$H$^+$ \citep[][]{Deguchi90,Zhang08}, and the first-ever astrophysical detection of HeH$^+$ \citep[][]{Gusten19}. 
Radio interferometric maps of NGC 7027 in emission from HCO$^+$ \citep{Deguchi92, Likkel92, Huang10} , CO \citep{Bieging91, Fong06}, and $^{13}$CO \citep{Nakashima10} have previously been obtained, typically at 2-4$''$ resolution. 
A recent, comprehensive HST imaging study from near-UV to near-IR (Figure \ref{fig_7027HST}) provides additional detail concerning this collimated outflow system, including evidence for strong shocks along one of the three main outflow axes \citep{Moraga2022}. However, since the aforementioned interferometric observations, made between 2008 and 2010 \citep{Zhang08, Huang10, Nakashima10}, no radio-based mapping of NGC 7027 has been published. 

NGC 7027 also constitutes a well-studied example of UV-excited fluorescent emission from molecular hydrogen, which arises from within a photodissociation (or photon-dominated) region (PDR) along the inner boundary of its massive molecular envelope \citep{Graham93}.
\cite{Latter00} established the detailed structure of this hot H$_2$-emitting region via near-IR imaging with the 
Hubble Space Telescope.
Followup ground-based near-IR imaging spectroscopy of H$_2$ was interpreted as revealing multiple collimated outflows scattered over a wide range of outflow axes \citep{Cox02}, providing the basis for the multipolar outflow model later proposed by \cite{Nakashima10}. 

In this paper, we present interferometric mapping of NGC 7027 in two highly reactive molecular ions, CO$^+$ and HCO$^+$. These two molecules are particularly interesting for purposes of understanding the structures and molecular chemistries of PDRs as well as X-ray-dominated regions (XDRs).
However, present theory leaves open the precise mechanisms whereby CO$^+$ and HCO$^+$ might trace PDR versus XDR chemical processes, and how the two molecules might act more generally as probes of non-LTE chemistry in astrophysical sources. This uncertainty serves as the prime motivation for our comparative observational study of CO$^+$ and HCO$^+$ emission from NGC 7027.

Enhanced abundances of CO$^+$ are likely found exclusively in such regions of high far-UV or X-ray flux.
In (O-rich) ISM PDRs, CO$^+$ can be formed via reactions between OH and C$^+$ \citep{Latter93}, where the C$^+$ is generated via photodissociation and then ionization of CO. If so, then CO$^+$ should primarily reside along PDR transition layers, between fully atomic and fully molecular gas. 

The sole empirical observation that supports such a connection between CO$^+$ and dense PDR environments is a single-dish map of CO$^+$ emission within the star-forming region Mon R2 \citep{TM16}. These observations established that CO$^+$ is cospatial with the PAH and [C {\sc ii}] emission found at the H{\sc i}/H$_2$ interface, as expected.

However, in the case of irradiation of molecular gas by photons with energies $>$14~eV, CO can also directly form CO$^+$. Furthermore,
\cite{Spaans07} have shown that significant X-ray flux 
will boost CO$^+$ abundance through enhanced production of C$^+$ and OH. Because X-rays penetrate higher column densities of molecular gas more readily than UV photons, X-ray irradiation should enable CO$^+$ production beyond the reach of PDRs, i.e., within the surrounding XDRs.

Primary destruction paths of CO$^+$ consist of dissociative recombination as well as reactions with H$_2$ or with neutral H at intermediate depths \citep[e.g.,][]{Stauber09}. One of the possible CO$^+$ destruction products is HCO$^+$, via the reaction
\begin{equation}
\mathrm{CO^+} + \mathrm{H_2/H} \rightarrow \mathrm{HCO^+/HOC^+/CO} + \mathrm{H}.
\end{equation}
Modeling by \cite{Bell07} shows CO$^+$ line intensity to increase weakly in conjunction with that of HCO$^+$, but were unable to draw conclusions as to whether the presence of CO$^+$ leads to HCO$^+$ enhancement. 
If such were the case, one would expect to see cospatial HCO$^+$ and CO$^+$ emission.

Alternatively, HCO$^+$ will form at large depths in X-ray-irradiated molecular gas via
\begin{equation}
\mathrm{H^+_3} + \mathrm{CO} \rightarrow \mathrm{HCO^+} + \mathrm{H_2},
\end{equation}
where H$^+_3$ abundance enhancement is due to X-ray ionization of H$_2$ \citep[e.g.,][]{Deguchi90}. Models of gaseous nebulae have indeed shown that the strong X-ray irradiation characteristic of XDRs is required to produce observed column densities of HCO$^+$ \citep{Kimura12}. In contrast,  photodissociative recombination of H$^+_3$ with electrons should limit HCO$^+$ production in PDR environments. 

The foregoing implies that direct comparison of the CO$^+$ and HCO$^+$ spatial distributions in planetary nebulae should provide a straightforward test of XDR- vs PDR-like chemistry. Specifically, if FUV/EUV photons dominate CO$^+$ production, in a PDR (through C$^+$~+~OH), then one expects CO$^+$ to trace the C$^+$ layer \citep[as in Mon R2;][]{TM16}, which should in turn coincide with a region of fluorescent (near-IR) H$_2$ emission \citep{Graham93}. HCO$^+$ would then either be found in an XDR in the surrounding molecular envelope, where 
CO~+~H$_3^+$ becomes an efficient HCO$^+$ production mechanism, or in the PDR, if dissociative recombination of CO$^+$ drives HCO$^+$ production. A third possibility is that CO$^+$ and HCO$^+$ both derive from CO, through direct X-ray ionization and via collisions with H$_3^+$, respectively; in this case the two species would also be expected to coincide spatially, but in an XDR that surrounds the PDR traced by near-IR H$_2$.

To date, HCO$^+$ has been detected in a few dozen planetary nebulae \citep[e.g.,][]{Schmidt16,Schmidt17,Bublitz19}, and (as noted) has been mapped in NGC 7027, while CO$^+$ emission has been identified in seven PNe \citep{Latter93, Bell07}. However, the detailed spatial distribution of the latter molecule remains to be established in any PN.  To fill this void, we have conducted an interferometric mapping study of NGC 7027 in CO$^+$ $J = 2\rightarrow1$ and HCO$^+$ $J = 1\rightarrow0$ emission with the Institut de Radioastronome Millim\'{e}trique's (IRAM) NOrthern Extended Millimeter Array (NOEMA). 

In Section 2, we describe the NOEMA and supplementary (archival) observations. The results, including CO$^+$ and HCO$^+$ velocity channel maps and velocity-integrated images, are presented in Section 3. Section 4 presents an analysis of the structures of the CO$^+$ and HCO$^+$ emitting regions in NGC 7027, including a comparison with each other and with archival CO and H$_2$ imaging, in addition to new evidence of the nebula's multi-polar jet system as traced by HCO$^+$ emission. Conclusions are presented in Section 5.

\section{Observations}

\subsection{NOEMA Data}

Observations of NGC 7027 were carried out at the $J = 2\rightarrow1$ rotational transition of CO$^+$ (235.380 GHz) and the $J = 1\rightarrow0$ transition of HCO$^+$ (89.189 GHz) and adjacent continuum with the Northern Extended Millimeter Array (NOEMA) interferometer in 2017 March-May and 2018 April, respectively. 
A phase center of the nebula R.A.(2000.0) = 21$^h$07$^m$01$^s$.8, dec.(2000.0) = 42$\degree$14$'$10$''$ was chosen for the CSPN \citep{Huang10}. 
All NOEMA antennae are equipped with dual polarization, heterodyne receivers that provide a bandwidth of 7.722 GHz across the lower and upper sidebands. 
The wide-band correlator WideX was configured for a contiguous bandwidth of 3.6 GHz and channel spacings of 1.95 MHz, for a velocity resolution of 2.48 km s$^{-1}$ at the CO$^+$ line. Eight antennae were used in the most compact configuration D and an on-source time of 9.6 hours. Baselines ranged from 24--176~m, with an angular resolution of 1\dotarcsec51$\times$1\dotarcsec42 (P.A.=$-74\degree$) at 236 GHz and peak S/N in the continuum of $\sim$24.

The introduction in late 2017 of an additional antenna and the new wide-band PolyFiX correlator allowed us to achieve a velocity resolution of 0.5 km s$^{-1}$ for the HCO$^+$ line. The survey observations were performed using nine antennae in the intermediate configuration C, for an on-source time of 7.4 hours. Baselines of 24--704~m achieved an angular resolution of 2\dotarcsec09$\times$2\dotarcsec07 (P.A. = 62$\degree$) for the 89.0 GHz HCO$^+$ line and peak S/N in the continuum of 36$\sigma$.   
A summary of observing details can be found in Table \ref{Line_Details}.

\begin{table}[htp]
\caption{\sc NOEMA Spectral Line Mapping of NGC 7027: Observation Summary}
\begin{center}
\begin{tabular}{l c c}
\hline
\hline
Molecule			& CO$^+$				& HCO$^+$ \\
\hline
Frequency	    	& 235.380 GHz			& 89.189 GHz \\
Transition			& $J = 2\rightarrow1$	& $J = 1\rightarrow0$\\
Baselines			& 24--176~m			    & 24--704~m \\
Beam Size	    	& 1.51${''} \times$1.42${''}$ & 2.09${''}\times$2.07${''}$ \\
Spectral Resolution	& 1.95 MHz			    & 0.149 MHz\\
Velocity Resolution	& 2.48 km s$^{-1}$		& 0.5 km s$^{-1}$ \\
S/N for Continuum	& 24$\sigma$			& 36$\sigma$\\
\hline
\end{tabular}
\end{center}
\label{Line_Details}
\end{table}%

Phase and amplitude calibrations utilized 2200+420, J2050+363, and J2120+445. Bandpass calibrations were performed with 3C84, 3C454.3, J2120+445, or 2013+370, with MWC349 for absolute flux calibration (1.08 Jy at 89 GHz, 1.94 Jy at 230 GHz). 
For the centered pointing, the phase center was used, while pointing and focusing errors of less than 30\% were reported during observations. Typical seeing varied day-to-day, with an average of 0\dotarcsec84 for CO$^+$ and 0\dotarcsec65 for HCO$^+$.

\begin{figure*}
	\includegraphics[width=\textwidth]{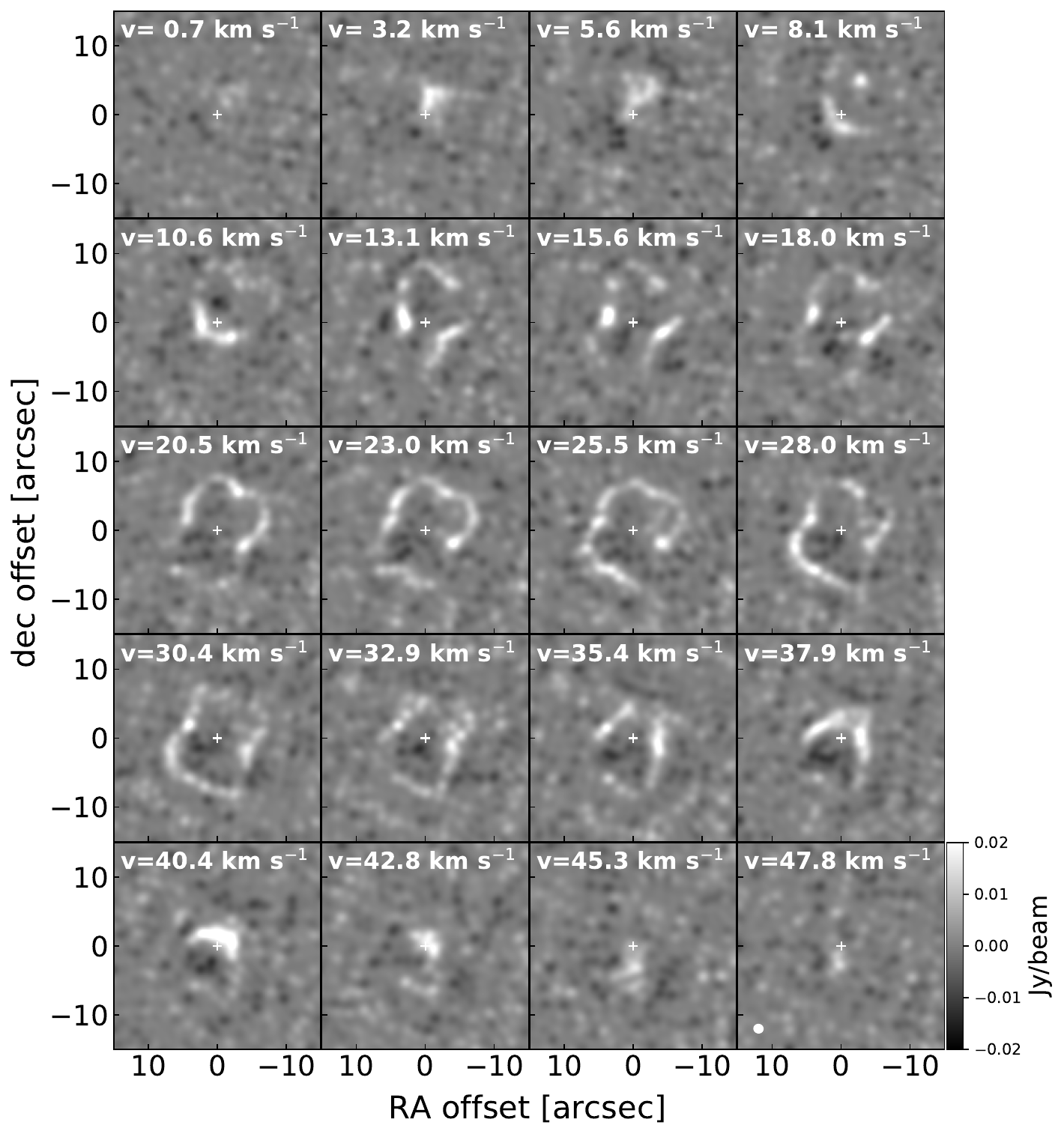}
	\caption{NOEMA channel maps of CO$^+$(2--1) emission from NGC 7027, displayed at $\sim$2.5 km s$^{-1}$ intervals. Flux intensity scale is at lower right. Channel LSR velocity is indicated in each frame. The beam size is indicated by the white oval in the lower right frame and crosses indicates the position of the central star.}
	\label{fig_COchannelmap}
\end{figure*}

\begin{figure*}
	\includegraphics[width=\textwidth]{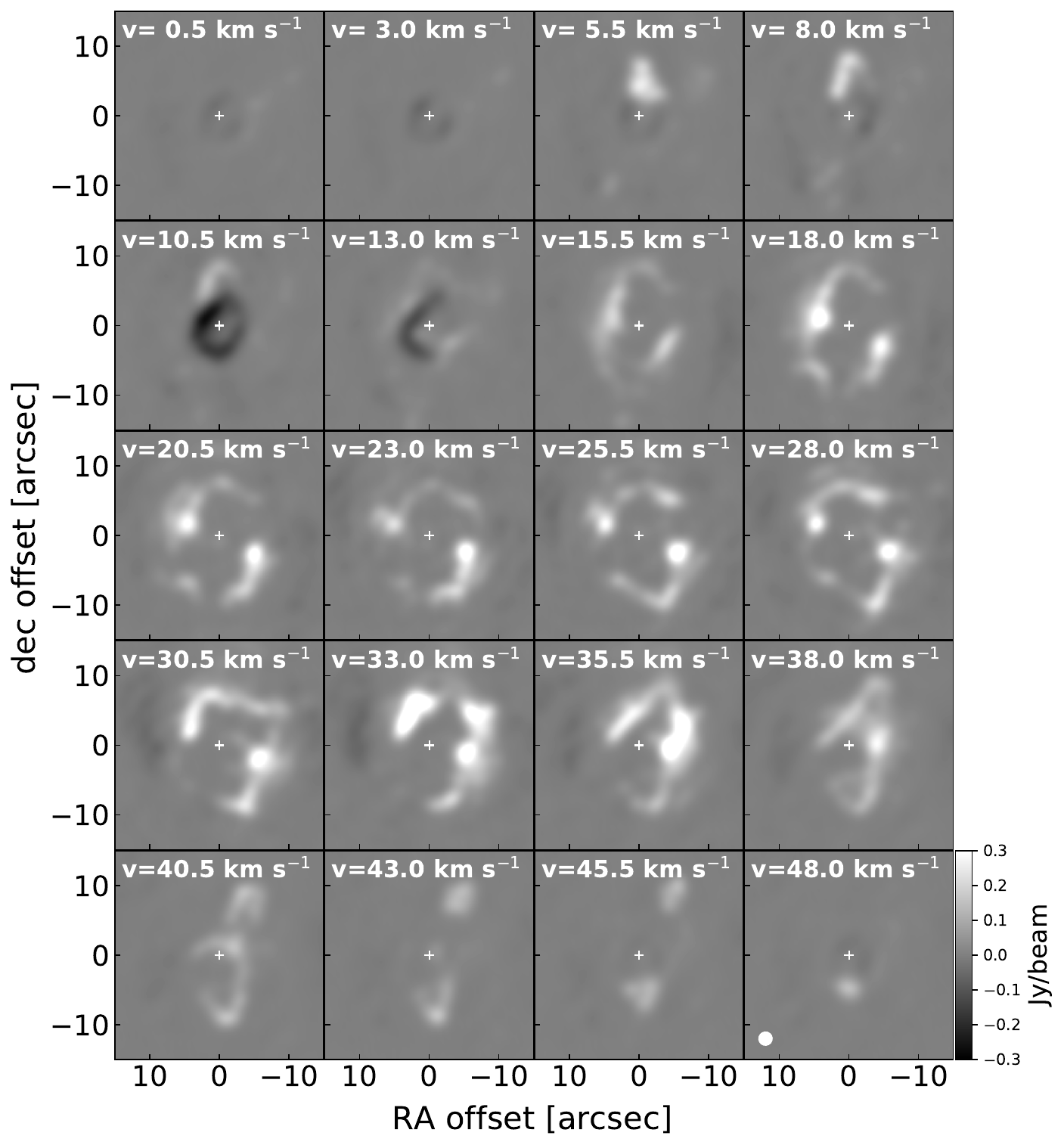}
	\caption{NOEMA channel maps of HCO$^+$(1--0) emission from NGC 7027, presented at the same (2.5 km s$^{-1}$) velocity intervals as for CO$^+$(2--1) (Figure \ref{fig_COchannelmap}). Flux intensity scale is at lower right. Channel LSR velocity is indicated in each frame. The beam size is indicated by the white oval in the lower right frame and crosses indicates the position of the central star.} 
	\label{fig_HCOchannelmap}
\end{figure*}

\begin{figure*}
	\centering
	\includegraphics[width=0.49\textwidth]{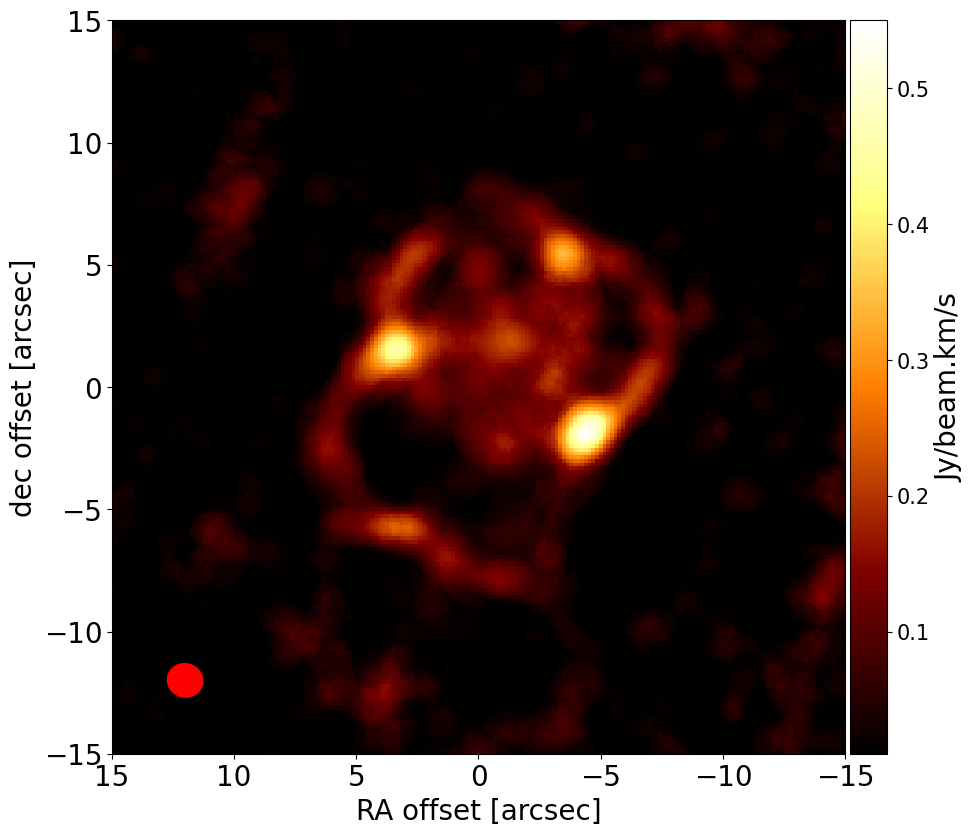}
	\includegraphics[width=0.49\textwidth]{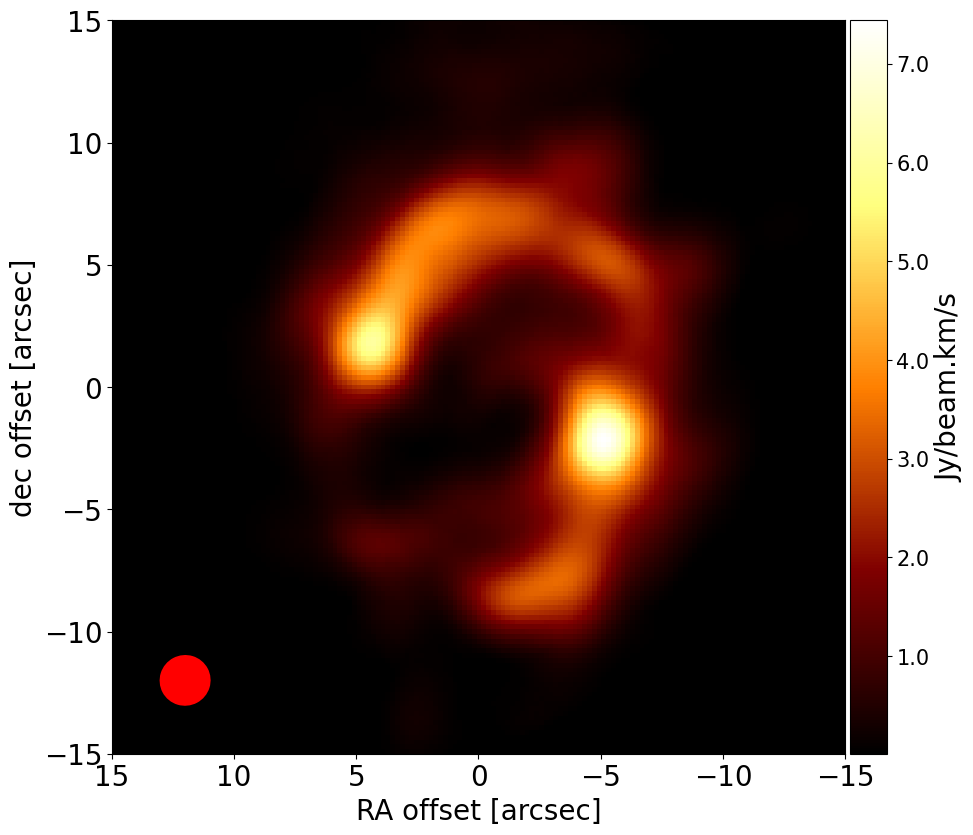} \\
	\includegraphics[width=0.48\textwidth]{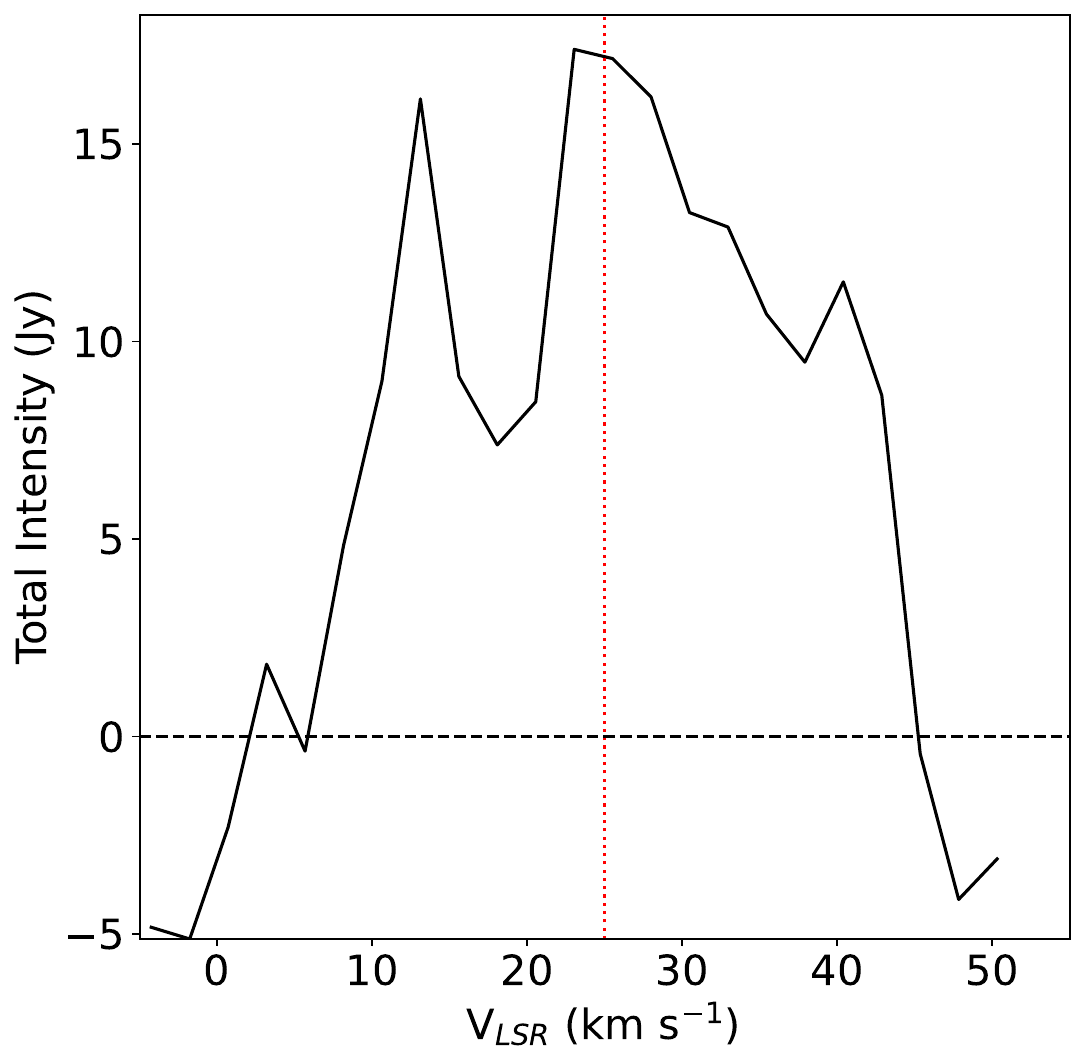}
	\includegraphics[width=0.50\textwidth]{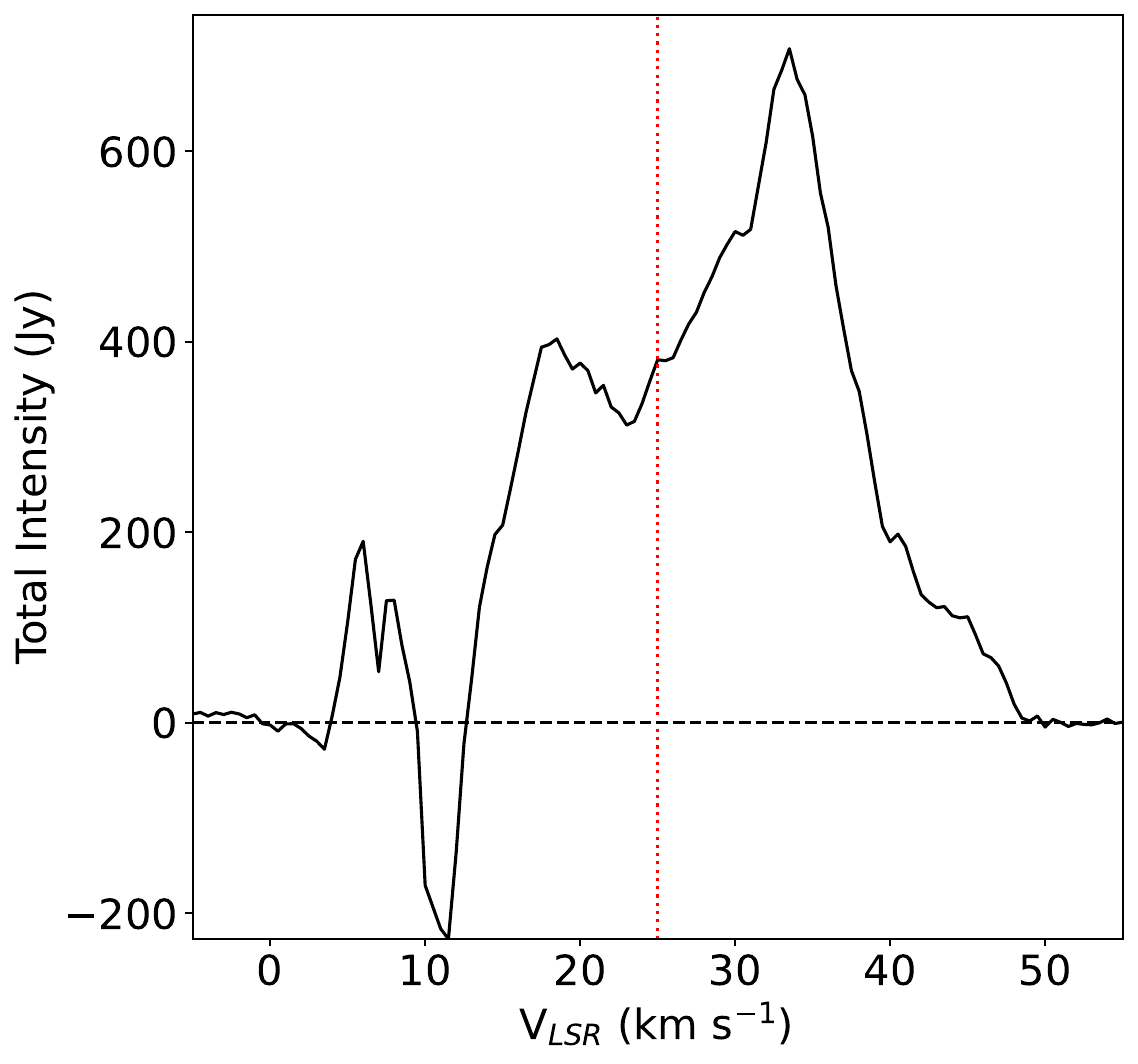}
	\caption{\textit{Top:} CO$^+$(2--1) (left) and HCO$^+$(1--0) (right) emission integrated over the full range of channels that display signal (0.7--47.8 km s$^{-1}$ for CO$^+$, 4.5--50 km s$^{-1}$ for HCO$^+$). In each frame, North is up, East is left, and the beams are represented by red ovals in the lower left corners.
	\textit{Bottom:} Spectra of CO$^+$(2--1) (left) and HCO$^+$(1--0) (right) emission, obtained as the spatially integrated intensities in the channel maps. The systemic velocity of NGC 7027 ($V_{LSR} =$ 24.3 km s$^{-1}$) is indicated by the dotted line.} 
	\label{fig_Mom0}
\end{figure*}

Data calibration was performed at the IRAM headquarters in Grenoble, France with the Continuum and Line Interferometer Calibration (CLIC) software package\footnote{http://www.iram.fr/IRAMFR/GILDAS/}.  
A standard calibration script was followed for radiometric corrections, as well as phase, flux, and amplitude calibration to produce the UV tables. 
Continuum subtraction and self calibration were also performed, whereby strong continuum emission was used to calibrate the adjacent line data. 
Natural weighting was used to generate images from the visibilities, and then deconvolution was performed with the Hogbom clean algorithm. 
The resulting high-quality velocity channel maps of NGC 7027 in CO$^+$(2--1) and HCO$^+$(1--0) are presented in Figures \ref{fig_COchannelmap} and \ref{fig_HCOchannelmap}, respectively; velocity-integrated (``moment 0'') CO$^+$(2--1) and HCO$^+$(1--0) images, as well as spectra extracted from the channel maps, are presented in Figure \ref{fig_Mom0}.


\subsection{SMA and HST data}

To place our NOEMA CO$^+$ and HCO$^+$ imaging results in context, we make use of (radio) CO and (near-IR) H$_2$ imaging obtained with the Submillimeter Array (SMA) and HST/NICMOS, respectively.
SMA CO imaging of the nebula was obtained in the J=3$\rightarrow$2 transition (345.796 GHz) on August 28, 2016 as part of a spectral line survey of NGC 7027 (Patel et al., in preparation).  The array utilized the subcompact configuration, with baseline lengths of 9.5--69.1~m, and a synthesized beam size of 3$''\times$2$''$. As such it should be noted that while the CO map adequately identifies the borders of the bulk of CO-rich gas in the nebula, it does not necessarily represent the total extent of CO. 
The observations were carried out in single pointing mode, with 15$''$ as the largest angular size of a continuous structure imaged by the subcompact array at this frequency. The phase center coordinates were R.A.(2000.0) = 21$^h$07$^m$01$^s$.5, dec.(2000.0) = 42$\degree$14$'$09.9$''$, while also used as the pointing center, and source velocity v$_{LSR}$ of 25 km s$^{-1}$. All 8 antennae available at that time were operational, and the DSB T$_{sys}$ values ranged from 400 to 600 K, varying between antennae and with source elevation. The zenith optical depth at 225 GHz was 0.1 throughout the track.  The absolute flux calibration was determined by observations of Titan and Neptune. Complex visibility gain calibration was performed by observing BL Lac and MWC 349a periodically for 3 min each, between the observations of NGC 7027 for 12 min. The bright quasars 3C454.3 and 3C279 were observed for 1.5 hours each, for spectral bandpass calibration.
The visibilities were calibrated using the MIR package in IDL\footnote{https://lweb.cfa.harvard.edu/rtdc/SMAdata/process/overview/}. Imaging was performed using the $clean$ task in CASA\footnote{https://casa.nrao.edu/}. 
Continuum emission was subtracted prior to imaging the spectral line emission, using the $uvcontsub$ task, identifying line-free channels in the u-v spectra by visual inspection. These data are among the first observations carried out with the SWARM correlator \citep{Primiani16}, with a spectral resolution of 140 KHz per channel, which was smoothed by a factor of 8 to obtain higher S/N with a spectral resolution of $\sim$0.97 km s$^{-1}$. The rms noise level in the imaged data cube is $\sim$150 mJy/beam.

A map of 2.12 $\mu$m rovibrational emission from molecular hydrogen was generated from HST/NICMOS images available in the Hubble Legacy Archive \cite[Project 7365, PI: William Latter;][]{Latter00}. The 2.12 $\mu$m H$_2$ emission was isolated by appropriately weighting and subtracting images obtained using the NICMOS F190N and F215N filters from the F212N filter image, to remove contaminating continuum and He {\sc i} emission, respectively.

\section{Results}

Figure~\ref{fig_COchannelmap} constitutes the first interferometric CO$^+$ mapping of NGC 7027, or of any PN. Indeed, to our knowledge, this is only the second CO$^+$ map obtained for any astrophysical source after that of the Mon R2 region \citep{TM16}. Emission from the $N = 2\rightarrow1$ 236.1 GHz CO$^+$ line spans roughly 1--48 km s$^{-1}$, with a central velocity of $V_{LSR} =$ 24.3 km s$^{-1}$ (Figure \ref{fig_COchannelmap}), and a total CO$^+$ flux of 391.2 Jy~km~s$^{-1}$. The nebula displays a multi-lobed structure in CO$^+$, as seen most clearly near the systemic velocity and in the velocity-integrated moment 0 image (Figure \ref{fig_Mom0}). The channel maps demonstrate that, with the exception of the nebular waist (see below), the most blue-shifted CO$^+$ is largely confined in the northeast regions of the nebula, while the most red-shifted CO$^+$ is found in its southwest regions.
As is discussed in more detail below, the overall CO$^+$ emission morphology and kinematics, as revealed in both the channel maps (Figure \ref{fig_COchannelmap}) and the velocity-integrated image (Figure \ref{fig_Mom0}), are strikingly similar to those of near-IR (2.12 $\mu$m) H$_2$ emission \citep[][their Figures 5, 6]{Cox02}. The brightest CO$^+$ emission is detected along the waist or minor axis of the nebula (where the major axis is oriented along position angle $P.A.=-25\degree$). Integrated over velocity (Figure \ref{fig_Mom0}), the CO$^+$ emission is elliptical yet appears as an almond-shaped waist due to the view at an inclination of 35$\degree$ \citep[see also][]{Cox02}.

\begin{figure*}[htbp]
	\begin{center}
	\includegraphics[width=0.49\textwidth]{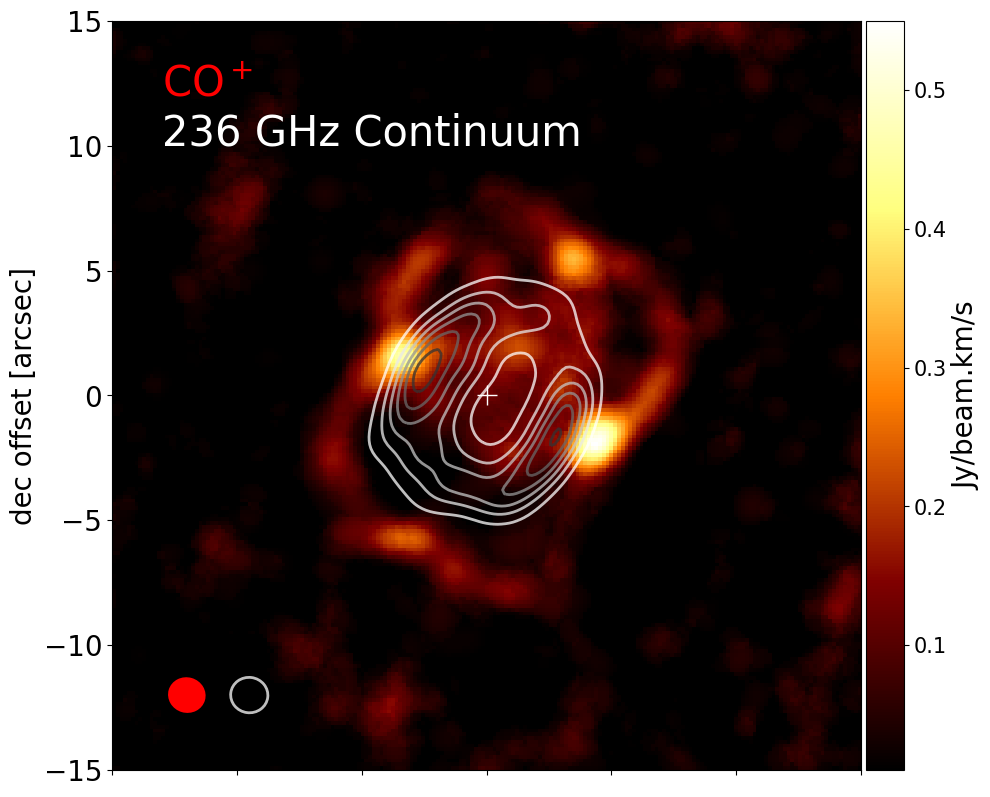}
	\includegraphics[width=0.49\textwidth]{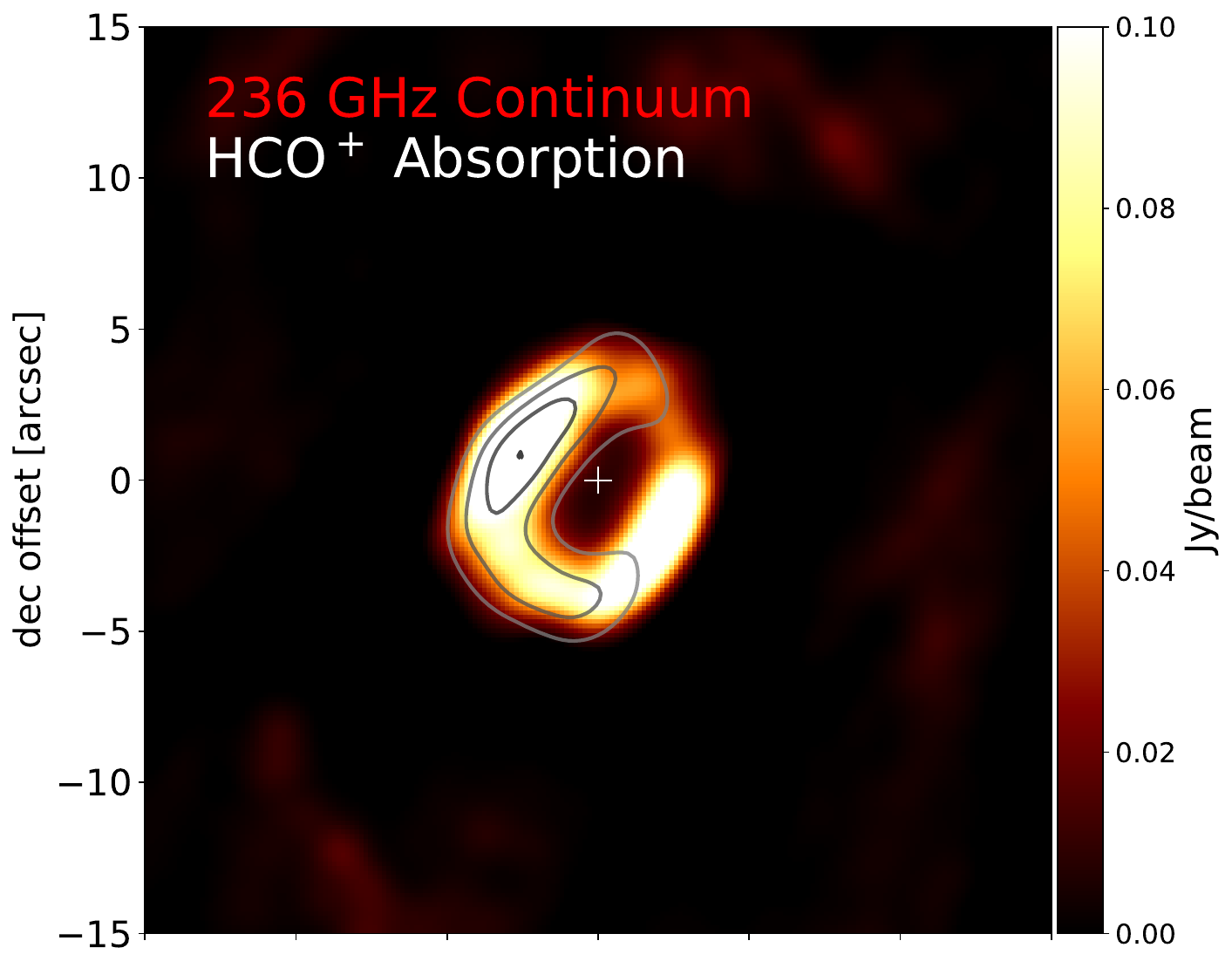}
	
	\includegraphics[width=0.49\textwidth]{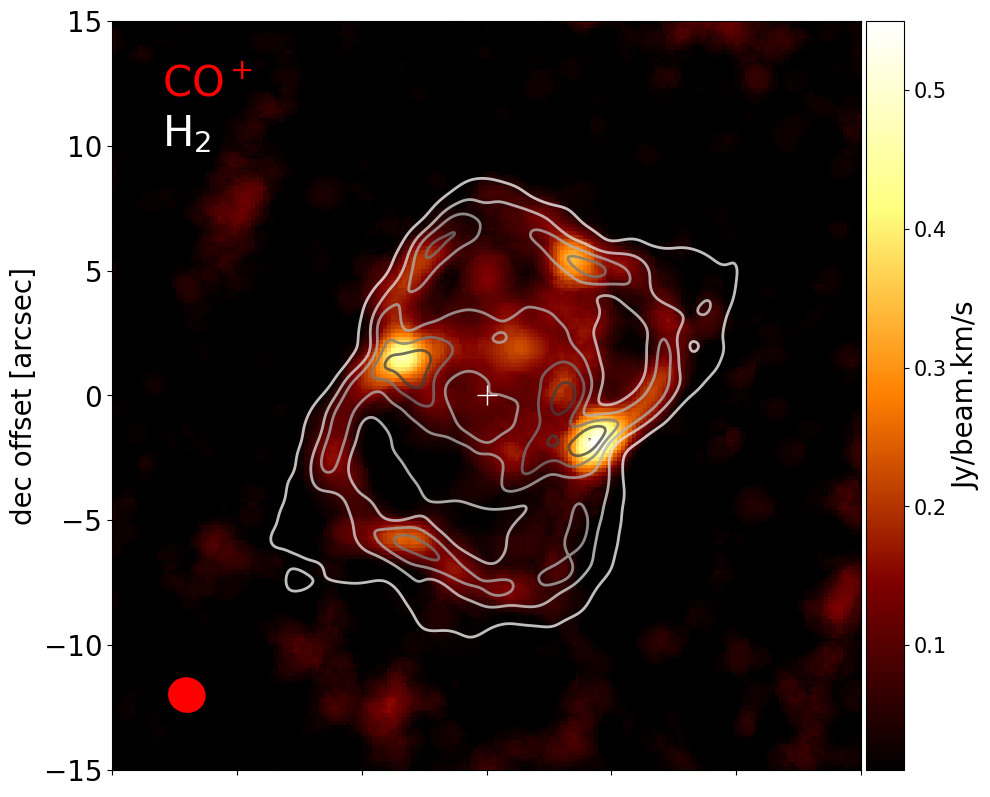}
	\includegraphics[width=0.49\textwidth]{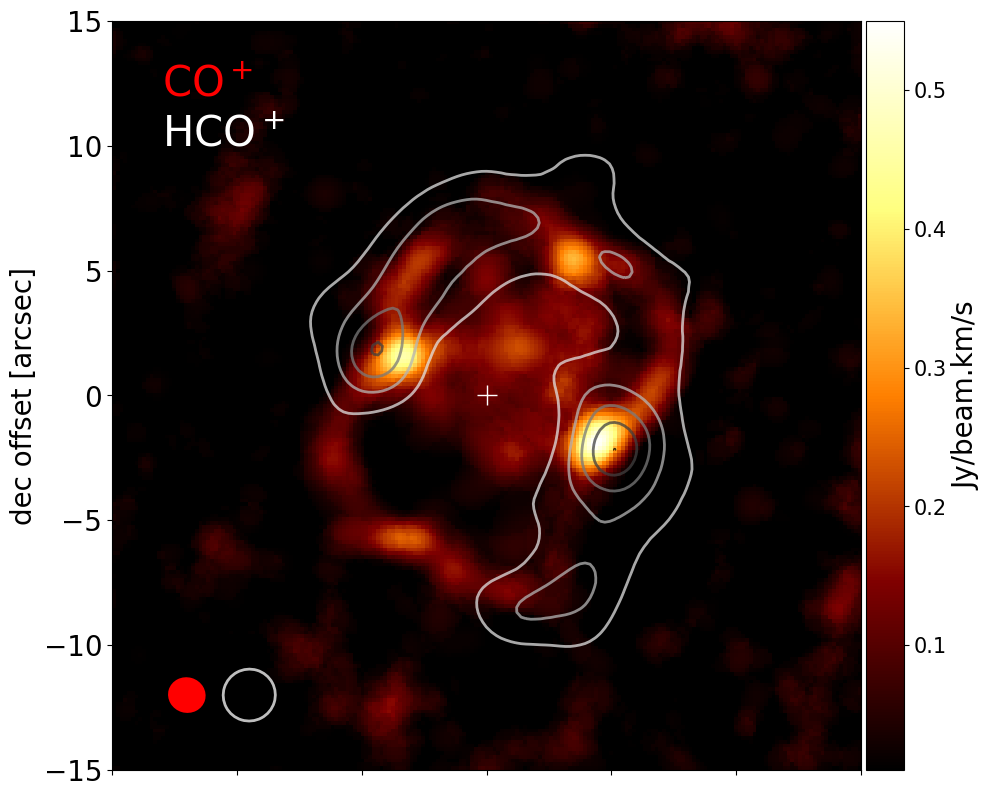}

	\includegraphics[width=0.49\textwidth]{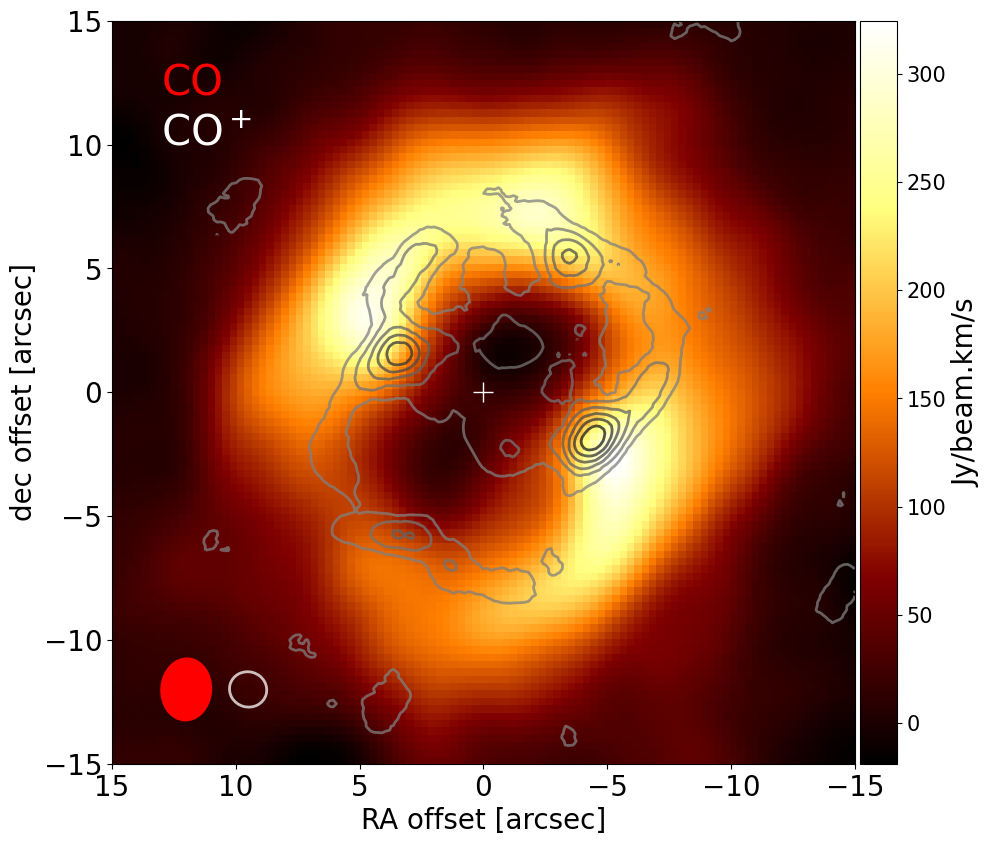}
	\includegraphics[width=0.49\textwidth]{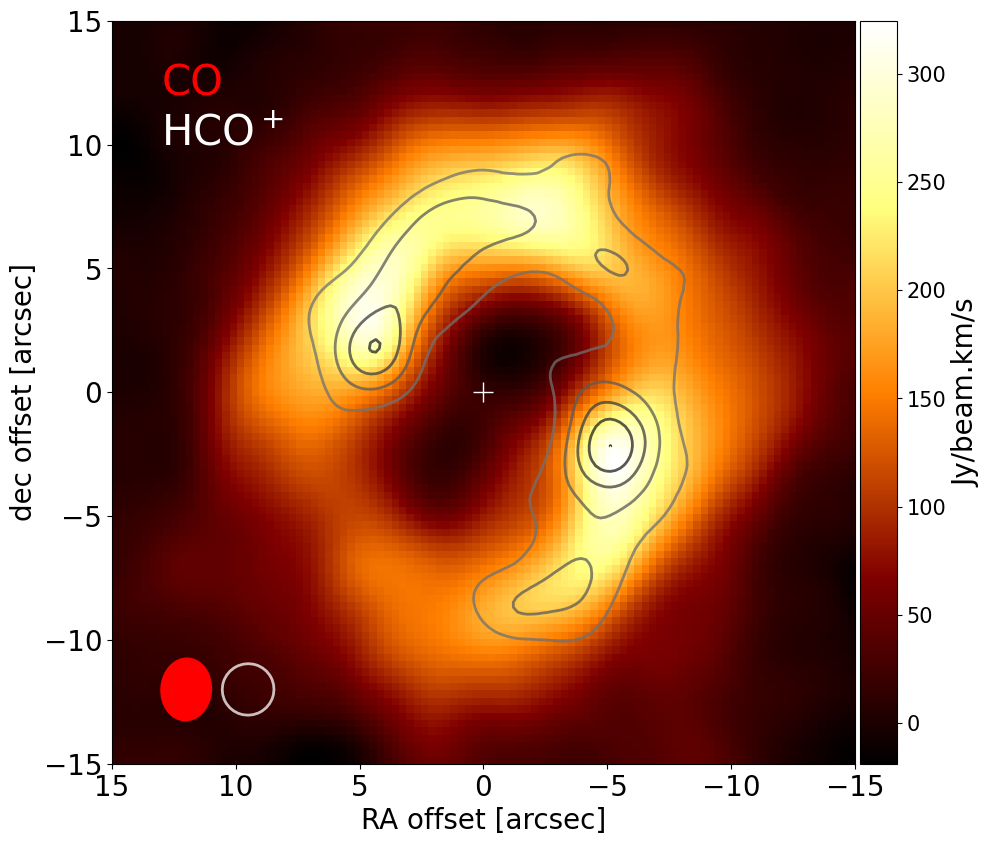}

	\caption{
	\textit{Top Left:} Contours of the 236 GHz continuum emission from NGC 7027 [0.02, 0.05, 0.08, 0.11, 0.14, 0.17, 0.20 Jy beam$^{-1}$] overlaid on a velocity-integrated CO$^+$ image. 
	\textit{Top Right:} Continuum extracted from off-channel lines in CO$^+$ observations. Overlaid contours denote the absorption feature seen from 8.5--13.5 km/s in the velocity averaged image of HCO$^+$ [$-0.2, -0.15, -0.10, -0.5$ Jy beam$^{-1}$ km s$^{-1}$ ].
	\textit{Middle:} Molecular map of CO$^+$ overlaid with contours of H$_2$ (left, smoothed to the CO$^+$ beamsize) [0.00, 2.24, 4.48, 6.72, 8.96, 11.20 Jy beam$^{-1}$ km s$^{-1}$ ] and of HCO$^+$ (right) [0.00, 1.49, 2.98, 4.46, 5.95, 7.44 Jy beam$^{-1}$ km s$^{-1}$ ].
	\textit{Bottom:} SMA CO (J=3--2) image of NGC 7027 (Patel et al., in preparation) overlaid with contours of CO$^+$ (left) [0.08 0.16 0.24 0.32 0.4  0.48 0.56 Jy beam$^{-1}$ km s$^{-1}$ ] and HCO$^+$ (right, same as contours above).
	Beam sizes for color plots (filled red) and contours (white outline) are included in the lower left of each image.
	} 
    \label{fig_comparison}
	\end{center}
\end{figure*}

Continuum emission in the vicinity of the CO$^+$(2--1) line (mean frequency 236.5 GHz; Figure \ref{fig_comparison}, top left) is well detected, and appears as an elliptical region whose major axis is well aligned with that of the velocity-integrated CO$^+$ emission (i.e., P.A $\sim -30\degree$). The peak continuum emission is seen along the waist (minor axis) of the nebula, and there is a prominent gap or hole in the elliptical emitting region $\sim$4-5$''$ northwest of the ellipse's center (i.e., the position of the central star). The dimensions of the 1.3 mm continuum emission are 11.6${''} \times$9.1${''}$, consistent with previous mm-wave mapping studies \citep[e.g.,][and references therein]{Huang10}.

The 89.6 GHz HCO$^+$(1--0) line has here been mapped by NOEMA at an unmatched $\sim$2${''}$ resolution (Figures \ref{fig_HCOchannelmap}, \ref{fig_Mom0}). In Figure \ref{fig_HCOchannelmap}, the native PolyFiX correlator channel resolution of $\sim$0.5 km s$^{-1}$ has been rebinned to 2.5 km s$^{-1}$ to match that of CO$^+$(2--1).  The HCO$^+$ emission is observed to span a total velocity width of 47.5 km s$^{-1}$, with a central velocity of $V_{LSR} =$ 24.8 km s$^{-1}$. 
We measure a total HCO$^+$(1--0) flux of 1970 Jy km s$^{-1}$.
Emission from HCO$^+$ appears as an extended, evacuated shell, with peak emission again appearing along the nebular minor axis and a non-axisymmetric dearth of material along the SE quadrant (Figure \ref{fig_Mom0}). 
The HCO$^+$ kinematic signatures are overall similar to those seen in CO$^+$, with blue-shifted and red-shifted emission to the northwest and southeast, respectively, but additional structures are apparent at the more extreme blue-shifted and red-shifted velocities; in particular, prominent blue-shifted emission features appear in the northern regions of the nebula.

The line profile of HCO$^+$ extracted from velocity channels (Figure \ref{fig_Mom0}, lower right) displays a broad pedestal of emission over the range $\sim$5--50~km s$^{-1}$ interrupted by a deep, narrow absorption feature at $\sim$10--13~km s$^{-1}$.
In the HCO$^+$ channel maps (Figure \ref{fig_HCOchannelmap}, upper left), the absorption is indeed most clearly seen in the 10.5 and 13.0 km s$^{-1}$ frames, with a depth of $\sim-0.4$ Jy beam$^{-1}$ compared to the continuum-subtracted baseline. Previous emission line surveys have identified blue-shifted absorption and proposed that a cold obscuring layer of HCO$^+$ gas is responsible \citep{Deguchi90, Bublitz19}. Figure \ref{fig_HCOchannelmap} marks the first interferometric map of such a region in NGC 7027, and demonstrates that the region of absorption appears as a ring-like feature cospatial with that of the continuum emission in the nebula (Figure \ref{fig_comparison}, top right). This suggests that the blue-shifted HCO$^+$ absorption feature is formed by a layer of cold molecular gas in front of the continuum source, which is presumably free-free radiation from a dense shell of photoionized gas. Furthermore, we speculate that this layer of cold HCO$^+$ is at least moderately opaque for the absorption feature to be present.

\section{Analysis and Discussion}

\begin{figure*}
	\centering
	\includegraphics[width=0.9\textwidth]{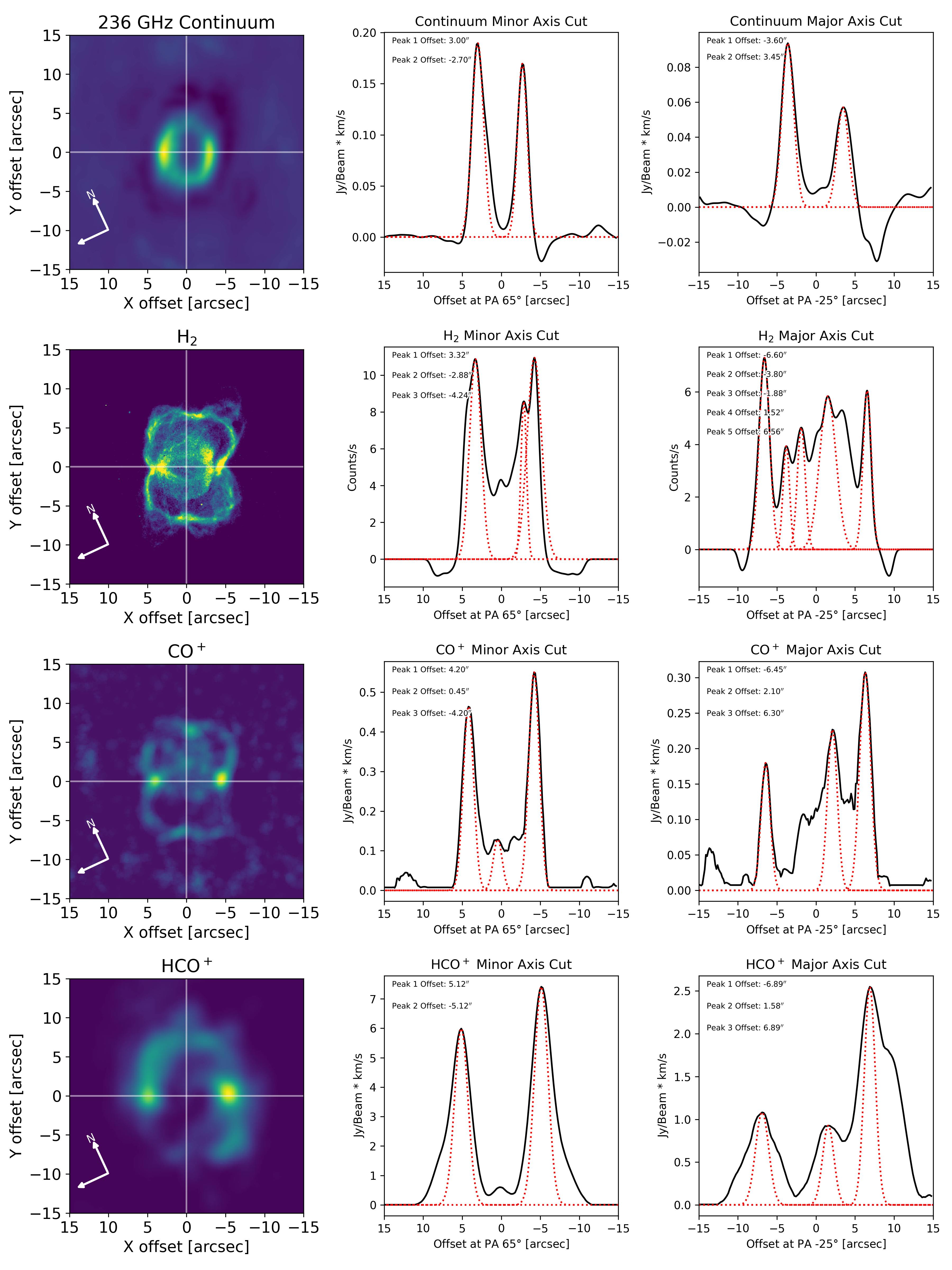}
	\caption{HST/NICMOS and NOEMA images of NGC 7027 with nebula major and minor axes marked in white (left column). Images are rotated -25$\degree$ with respect to their sky orientation to vertically align the major axis. N is marked in each figure. Intensity of emission along the axes are plotted on the right, with points of greatest emission marked by Gaussian fits in red. Offsets of these peaks in arcseconds from the center of the nebula are displayed in the upper left corner.} 
	\label{fig_1Dcuts}
\end{figure*}

Figure \ref{fig_comparison} presents a comparison of the spatial structure of velocity-integrated CO$^+$ and HCO$^+$ emission with the 1.3 mm NOEMA continuum image, in addition to the archival HST/NICMOS near-IR H$_2$ image and SMA CO(3--2) map. The left column compares the velocity-integrated CO$^+$ map with contours of continuum emission and H$_2$, as well as an overlay of CO$^+$ contours onto CO. The right column overlays contours of HCO$^+$ 
with the continuum, CO$^+$, and CO.
In Figure \ref{fig_1Dcuts} we display linear cuts through the moment 0 maps, along the nebula's major and minor axes ($P.A. = -25\degree$ and $65\degree$, respectively), through the images of CO$^+$ and HCO$^+$, as well as through the H$_2$ and the 1.3~mm continuum images. 
The linear cuts have been fitted with multiple Gaussian functions, to facilitate comparisons of the positions of emission peaks along the major and minor axes.
The discussions that follow in Sections \ref{sub_COp} and \ref{sub_HCOp} for CO$^+$ and HCO$^+$, respectively, are derived from the analysis presented in Figs.~\ref{fig_comparison} and \ref{fig_1Dcuts}. Section \ref{sec:HCOpJets} presents a kinematic analysis of the HCO$^+$ data.

\subsection{CO$^+$} \label{sub_COp}
Comparison of the NOEMA CO$^+$ image with that of the cleaned HST/NICMOS H$_2$ image reveals highly cospatial emission, in which both the ionized and neutral molecular gas trace the multi-lobed, pillow-like structure of NGC 7027 (Figure \ref{fig_comparison}, middle left). A prominent torus of material is present along the equatorial plane, with the brightest emission occurring at the two edges of the equatorial region. These intensity hot-spots also appear in the H$_2$ map, with the Eastern edge occurring marginally interior to CO$^+$. Similar emission peaks are detected at the NW and SE ends of the H$_2$ lobes, also coincident with CO$^+$ emission peaks. In contrast, the CO$^+$ and CO emission morphologies are strikingly different (Figure \ref{fig_comparison}, lower left). 

A minor axis cut across the H$_2$ map (i.e., along the nebula's equatorial plane) reveals a pair of emission peaks at the edges of the waist (Figure \ref{fig_1Dcuts}). Although the Eastern-most feature appears offset from the minor axis, it and its opposite component appear at 4.29$''$ and $-4.24''$ offset from the CSPN, while the inner bright components of the waist lie at 3.32$''$ and -2.88$''$. 
CO$^+$ displays somewhat narrower emission peaks along the minor axis at $\pm$4.20$''$ offsets, roughly aligning with the two outer H$_2$ peaks. The inner H$_2$ peaks do not have an apparent CO$^+$ counterpart, indicating CO$^+$ lies closer to the adjacent molecular rich regions. 
The emission along the major axis is similarly cospatial, with the outermost CO$^+$ peaks lying at a N-S average CSPN displacement of 6.4$''$ and the outermost H$_2$ at 6.6$''$, well within the respective ($\sim$0.8$''$) FWHMs of these local peaks in emission. 
It is hence apparent that, overall, the CO$^+$ and H$_2$ emission are tracing the same emitting region. 
In contrast, the 1.3~mm continuum emission has an extent of 2.85$''$ and 3.53$''$ along the nebular minor and major axes, respectively; assuming the continuum emission is due to free-free radiation from dense plasma, this indicates that the densest region of ionized gas lies well inside that of the CO$^+$ and H$_2$ emitting region.


As previously noted, 2.12 $\mu$m H$_2$ emission can serve as a UV irradiation tracer, and various studies have discussed the likelihood that, in NGC 7027, this H$_2$ rovibrational line is excited via UV fluorescence and, hence, traces a PDR within the nebula \citep{Graham93, Kastner96, Latter00, Cox02}. The strong similarity between the CO$^+$ and H$_2$ emission surfaces therefore serves as definitive evidence that, 
Although CO$^+$ has been flagged as a potential XDR tracer \citep{Wolfire22}, NGC 7027 appears to be a clear case where CO$^+$ traces the PDR.
Comparable results were found in Mon R2, where CO$^+$ resides at the atomic-molecular interface of the star-forming region \citep{TM16}. In that study, the spatial coincidence of CO$^+$ with PAHs and [C \textsc{ii}], which typically lie along the fluorescent H$_2$ emission region of PDRs, demonstrated that the CO$^+$ is UV-enhanced, via reactions with OH or perhaps CO charge transfer. Given the C-rich nature of NGC 7027, a sequence consisting of UV-driven H$_2$ photodissociation and subsequent photoionization of H, followed by charge transfer with CO, may be driving the formation of CO$^+$ in its PDR layer. 


\subsection{HCO$^+$} \label{sub_HCOp}
In comparing the HCO$^+$ total intensity map with that of CO$^+$ (Figure \ref{fig_comparison}, middle right), it is readily apparent that the emission morphologies of the two species diverge significantly. Minimal HCO$^+$ is present along the line of sight within the central cavity of NGC 7027, and the HCO$^+$ emission peaks, as well as the extended HCO$^+$ emission structure, lie significantly outside the CO$^+$ emission region. The linear cuts in Figure \ref{fig_1Dcuts} clearly illustrate the displacement of HCO$^+$ outside CO$^+$ along both the minor and major axes. Specifically, the bright HCO$^+$ peaks along the nebular waist lie 
$\sim$0.9$''$ outside their CO$^+$ counterparts, while the brightest HCO$^+$ emission along the major axis lies $\sim$0.5$''$ outside that of CO$^+$. It is also evident that detectable HCO$^+$ extends well beyond the outer CO$^+$ emission boundary.


Furthermore, in stark contrast to the general lack of correspondence of CO$^+$ and CO emission (Figure \ref{fig_comparison}, bottom left), HCO$^+$ closely follows the overall CO emission morphology (Figure \ref{fig_comparison}, bottom right). Emission from both molecules is similarly extended, with peak intensities near the nebular waist and weaker emission along a P.A. of 60$\degree$ (i.e., towards the SE quadrant) as well as toward the NW.

The foregoing strongly suggests that the HCO$^+$ abundance peaks in the extended molecular envelope of NGC 7027, well outside the region of peak CO$^+$ production. This in turn suggests that HCO$^+$ production proceeds through X-ray ionization of molecular gas in an XDR that lies outside of the PDR traced by CO$^+$ (and near-IR H$_2$). It remains to determine, via modeling, whether the source of X-ray ionization is the luminous, extended X-ray emission imaged by Chandra \citep{Montez18}, NGC 7027's exceedingly hot and luminous central star \citep{Moraga2022}, or some combination of the two. Regardless, production of HCO$^+$ via dissociative recombination of CO$^+$ in the nebula's PDR appears to be ruled out by these NOEMA observations.

\subsection{Multipolar Jets Traced by HCO$^+$}\label{sec:HCOpJets}

We consider here whether and how the CO$^+$ and (in particular) HCO$^+$ emission from NGC 7027 traces the complex system of outflows that appear to be impinging on and puncturing the inner, elliptical shell of ionized gas (seen in the 1.3~mm continuum image) as well as the nebula's outer dust ring system \citep[see][]{Moraga2022}.

\citet{Cox02} first raised the possibility of such a system of fast outflows in NGC 7027. They identified a number of cavities or breakouts, in the form of holes in their velocity-integrated near-IR H$_2$ emission map, which they attributed to multiple point-symmetric outflows that have pierced the otherwise conical bipolar shell that surrounds NGC 7027's ionized core. Tracing back from the holes suggests three sets of oppositely directed, narrow, collimated outflows (jets) emanating from the CSPN  (hereafter Outflows 1, 2, and 3\footnote{We have adopted the outflow naming convention of \citet{Cox02}, which corresponds to \citet{Nakashima10} directions A, B, and C as well as to their red, yellow, and blue bicones.}) have broken through the H$_2$ shell at some point in the nebula's recent history.
This interpretation is supported by subsequent modeling of CO emission-line interferometry by \citet{Nakashima10}, who estimated position angles of $-53\degree$, 4$\degree$, and $-28\degree$ and inclinations (with respect to the plane of the sky) of 55$\degree$, 35$\degree$, and $-25\degree$, respectively, for Outflows 1, 2, and 3.

As described in \S~\ref{sub_COp}, the velocity-integrated CO$^+$ map bears strong resemblance to the near-IR H$_2$ image --- including the precise positions of emission deficits and peaks --- such that the same (three) collimated outflow punctures seen in velocity-integrated H$_2$ are also present in velocity-integrated CO$^+$ \citep[compare our Figure \ref{fig_Mom0} with Figure 10 in][]{Cox02}. Furthermore, comparison of our Figure \ref{fig_COchannelmap} with Figure 6 in \citet{Cox02} demonstrates that the CO$^+$ kinematics closely match those of H$_2$.

\begin{figure*}[htbp]
	\begin{center}
	\includegraphics[width=\textwidth]{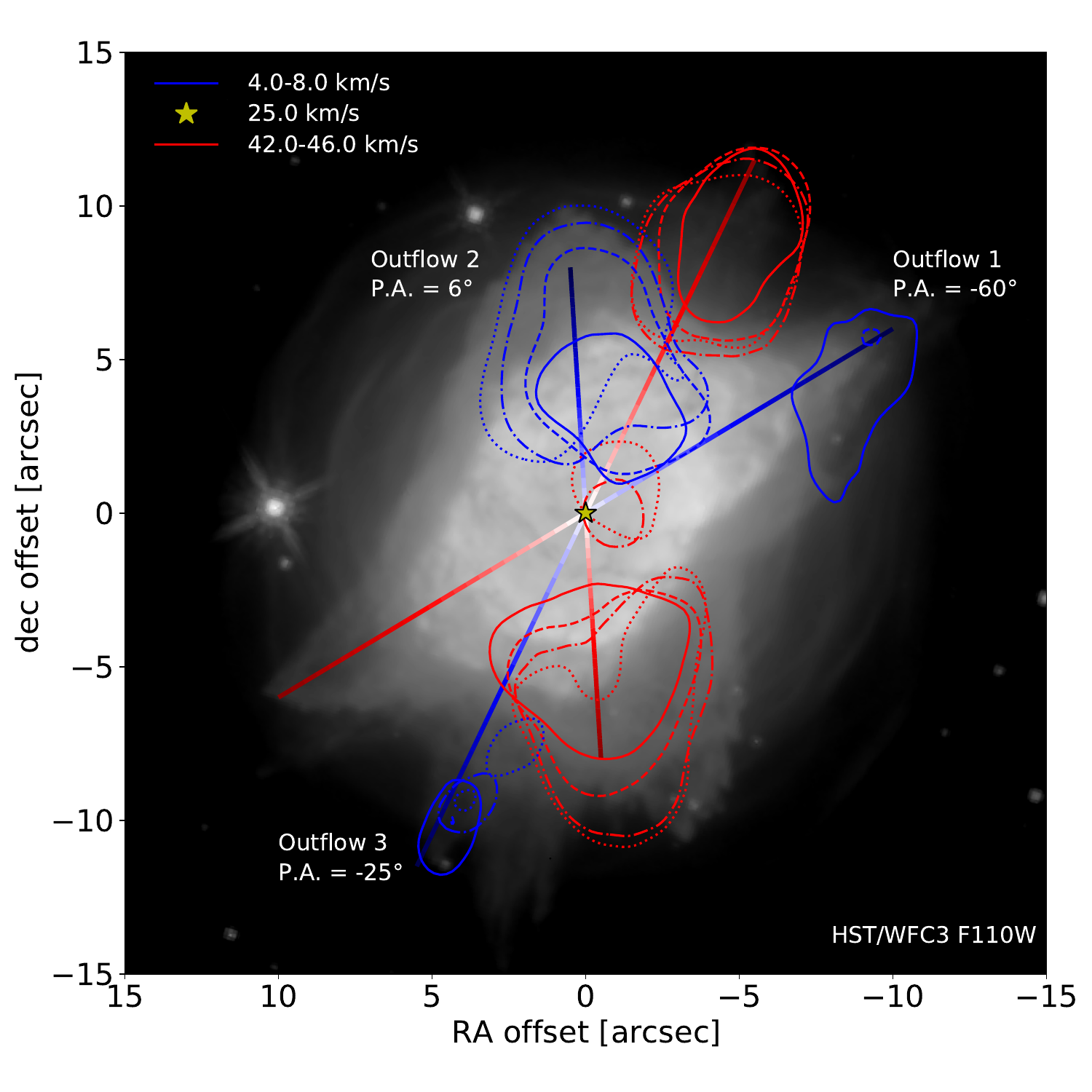}
	\caption{\textit{Left:} WFC3 F110W filter image of NGC 7027 with 15\% peak intensity contour overlays from HCO$^+$ of the 4--10 km s$^{-1}$ velocity channels (blue-shifted) and 40-46 km s$^{-1}$ velocity channels (red-shifted) expanding from the central velocity at 25 km s$^{-1}$. The HCO$^+$ emission traces the fast outflows from the CSPN and their orientation in the sky. Solid contours represent greatest velocity shift, while dashed, dash-dotted, and dotted contours denote velocities closer to the systemic velocity. Solid lines along position angles of 6$\degree$, $-25\degree$, and $-60\degree$ represent the approximate length and red- or blue-shifted orientation of the HCO$^+$ gas. }
	\label{fig_red-shiftblue-shift}
	\end{center}
\end{figure*}

Given emission from HCO$^+$ closely traces that of CO (Figure \ref{fig_comparison}), we might expect to see similar evidence for
the presence of multiple jets associated with the outflows, in the form of spatially localized high-velocity HCO$^+$ emission \citep[as described by][in the case of CO]{Nakashima10}.
Indeed, we have isolated three distinct regions of blue-shifted emission at $\sim$15--21 km s$^{-1}$ from the nebular systemic velocity, two of which have red-shifted counterparts at similar velocity offsets (Figure \ref{fig_red-shiftblue-shift}).
The strongest blue-/red-shifted emission pair lies at P.A. = 6$\degree$, along Outflow 2. 
The next strongest pair, along Outflow 3, lines up well with the major axis of NGC 7027 at P.A. = $-25\degree$, but appears inverted with respect to Outflow 2 due to its inclination, with red-shifted emission north of the central star and blue-shifted toward the south. As also discussed in \citet{Moraga2022}, this orientation suggests the outflow lies roughly along the plane of the nebular waist. 
Meanwhile, Outflow 1 only displays weak high-velocity blue-shifted HCO$^+$ emission at P.A. = $-60\degree$, toward the NW corner of the PN; no red-shifted emission counterpart is detected. 

The alignment of high-velocity Br$\gamma$ emission with Outflow 1 \citep[][]{Cox02} points toward it being the most recent occurrence of a processing, episodic outflow. This conclusion is supported by evidence that the strongest shocks lie along this same direction \citep{Moraga2022}. These shocks may be dissociating the molecular gas along the direction of Outflow 1, resulting in weak HCO$^+$ (and CO) emission relative to that observed along Outflows 2 and 3. 
The mechanism behind the asymmetric shock front, whereby HCO$^+$ gas in the SE is fully dissociated, yet HCO$^+$ in the NW remains intact, is as yet unclear.

\section{Conclusions}

With a massive molecular envelope that is exposed to intense UV and X-ray irradiation from its central star and associated wind shocks, the young and rapidly evolving PN NGC 7027 provides an unrivaled opportunity to study, and attempt to distinguish between, photon-dominated regions (PDRs) and X-ray-dominated regions (PDRs). In addition, while both CO$^+$ and HCO$^+$ are detected in various astrophysical environments, present theory leaves open questions as to how these two molecules form and whether they might serve as probes of PDR versus XDR chemistry. 
We have used the IRAM NOEMA interferometer to obtain maps of CO$^+$(2--1) and HCO$^+$(1--0) emission, at $\sim$2$''$ resolution, from NGC 7027. The former constitutes the first interferometric map of the molecular ion CO$^+$ in any astrophysical source.
The results reveal that CO$^+$ and HCO$^+$ trace fundamentally different nebular environments within NGC 7027. 
The CO$^+$ emission is precisely cospatial with near-IR H$_2$ emission, hence effectively tracing the inner boundaries of the PDR, where CSPN UV radiation drives the chemistry. In contrast, HCO$^+$ and neutral CO emission display similar morphologies, and comparison of the limb-brightened edges of the CO$^+$ and HCO$^+$ maps reveals a systematic radial offset, wherein regions of brightest HCO$^+$ lie $\sim$1$''$ ($\sim$900 au) outside of the corresponding CO$^+$ emission peaks. 
This indicates that HCO$^+$ is enhanced by dust-penetrating soft X-rays in an XDR that surrounds the PDR. High-velocity HCO$^+$ emission is also found to trace at least three distinct outflows from the central star at position angles of 6$\degree$, $-25\degree$, and $-60\degree$, providing additional evidence
that a rotating, episodic jet has repeatedly pierced the PDR in NGC 7027's short lifetime, generating X-ray-induced HCO$^+$ and, in the case of the most recent outflow (oriented SE-NW), fully dissociating shocks.
The comparison of interferometric CO$^+$ and HCO$^+$ maps of NGC 7027, in conjunction with near-IR H$_2$ and radio CO observations, thus serves to disentangle and spatially delineate the UV- and X-ray-induced PDR and XDR molecular environments within NGC 7027.

\begin{acknowledgments}
This work is based on observations carried out under projects W16BH and W17AV with the IRAM NOEMA Interferometer. IRAM is supported by INSU/CNRS (France), MPG (Germany) and IGN (Spain). 
We thank Michael Bremer (IRAM) for reduction of the NOEMA CO$^+$ data, and Charlène Lefèvre for support in the reduction procedure of the HCO$^+$ data. The Submillimeter Array is a joint project between the Smithsonian Astrophysical Observatory and the Academia Sinica Institute of Astronomy and Astrophysics and is funded by the Smithsonian Institution and the Academia Sinica. 
J.H.K. acknowledges the support of National Science Foundation grant AST-2206033 to Rochester Institute of Technology.
M.S.G, J.A, and V.B. are partially supported by the research grant Nebulaeweb/eVeNts (PID2019-105203GB-C21) of the Spanish AEI (MICIN).
I.A. acknowledges the support of CNPq, Conselho Nacional de Desenvolvimento Científico e Tecnológico - Brazil, process number 157806/2015-4 and the Coordena\c{c}\~{a}o de Aperfei\c{c}oamento de Pessoal de N\'{i}vel Superior - Brasil (CAPES) - Finance Code 001. 
The authors wish to thank Alex Wickham-Pirowski for initial development of NGC 7027 image rotation and intensity profile code, as well as the anonymous referee for helpful comments that improved the clarity of this paper.
\end{acknowledgments}

%

\vspace{5mm}
\facilities{NOEMA, SMA, HST}



\bibliography{LibraryOfAlexandria}{}
\bibliographystyle{aasjournal}



\end{document}